\documentclass[11pt,a4paper]{article}

 \usepackage[english]{babel}
 \usepackage[T2A]{fontenc}
 \usepackage[cp1251]{inputenc}
 \usepackage{amsmath}
\usepackage{graphicx}
\usepackage{amssymb}
\usepackage{color}
\usepackage{amsfonts}
\usepackage{wrapfig}
\usepackage{caption}

\newcommand{\no}{\nonumber}
\newcommand{\non}{\nonumber \\}

\newcommand{\be}{\begin{equation}}
\newcommand{\ee}{\end{equation}}
\newcommand{\bea}{\begin{eqnarray}}
\newcommand{\eea}{\end{eqnarray}}
\newcommand{\sli}{\sum\limits}
\newcommand{\ili}{\int\limits}
\newcommand{\lp}{\left (}
\newcommand{\rp}{\right )}

\newcommand{\vk}{\vec{k}}
\newcommand{\vl}{\vec{l}}
\newcommand{\vj}{\vec{j}}

\newcommand{\cB}{{\cal{B}}}

\newcommand{\rhok}{{\rho_{\vec{k}}}}
\newcommand{\rhomk}{{\rho_{-\vec{k}}}}

\begin{document}

\binoppenalty=10000
\relpenalty=10000

\begin{center}
\textbf{\Large{USING A CELL FLUID MODEL FOR DESCRIPTION OF A PHASE TRANSITION IN SIMPLE LIQUID ALKALI METALS}}
\end{center}

\vspace{0.3cm}

\begin{center}
M.P.~Kozlovskii, O.A.~Dobush\footnote{e-mail: dobush@icmp.lviv.ua} and I.V.~Pylyuk
\end{center}

\begin{center}
Institute
for Condensed Matter Physics of the National Academy of Sciences of
Ukraine \\ 1, Svientsitskii Str., 79011 Lviv, Ukraine
\end{center}

 \vspace{0.5cm}

\small This article embraces a theoretical description of the first order phase transition in liquid metals with application of a cell fluid model. The results are obtained through calculation of the grand partition function without usage of phenomenological parameters. The Morse potential is used for calculation of the equation of state and the coexistence curve. Specific results for sodium and potassium are obtained. Comparison of outcome of analytical expressions with data of computer simulations is presented.

\vspace{0.5cm}

PACs: 51.30.+i, 64.60.fd


\normalsize

\section{Introduction} \label{introduction}
This article is based on the method we proposed in~\cite{KD_2016}. It enable to obtain the equation of state of the cell model in wide range of temperatures below and above the critical point. Particular analytical results were conducted with use of the Morse potential
\begin{equation}\label{Pot_Morse_0}
    U(r) = \epsilon e^{-2(r-R_0)/\alpha} - 2 \epsilon e^{-(r-R_0)/\alpha}
\end{equation}
The consequence of the approach~\cite{KD_2016,KD_2017} is a restriction of the ratio between the coordinate of minimum $R_0$ and effective reach $\alpha$ of the interaction potential $R_0/\alpha <4 \ln{2}$. However according to numerical results~\cite{girifalko,lincoln} this ratio exceeds $R_0/\alpha = 4 \ln{2}$ for real substances, in particular, for fluid metals.
In present article the method proposed in~\cite{KD_2016}  and slightly changed in~\cite{KD_2017} is modified by means of introducing a temperature free effective interaction potential. This makes it applicable in the range $R_0/\alpha > 4\ln{2}$ for description of real metals in the region of a first order phase transition.

This paper is laid as follows:
in Section 2 the temperature free effective interaction potential is introduced and main steps of calculations towards obtaining an exact representation of the grand partition function of the cell fluid model are shown. This expression is restricted to $\rho^4$-model and calculated in the mean-field approximation in Section 3. Section 4 is dedicated to equation of state of the cell fluid applicable in wide temperature region except a vicinity of the critical point. In Section 4 the analytical result obtained in this manuscript is compared with simulation data~\cite{singh} for parameters of Morse potential describing alkali metals $Na$ and $K$. Discussion and conclusions are presented in Section 5.

\section{Representation of the grand partition function} \label{representation_GPF}

 The objective of our investigation is the description of behavior of a simple fluid in wide temperature region. For this purpose within the grand canonical ensemble we calculate the grand partition function (GPF) of the cell fluid model as an approximation of real continuous system and obtain the result in the form of a function of temperature and density.

The idea of the cell fluid [1,2] consist in fixed partition of the systems volume $V$, where $N$ particles reside, on $N_v$ mutually disjoint elementary cubes, each of volume $v = V/N_v$.
In a formalism of the cell model the GPF of a system of volume $V$ with $N$ particles is written in the form
\small
\be \label{GPF_1}
\small{ \Xi =\!\! \sli_{N=0}^{\infty} \! \frac{(z)^N}{N!} \!\! \int \limits_{V} \! (dx)^N \!
\exp \! \left[-\frac{\beta}{2} \!\! \sli_{\vec{j}_1,\vec{j}_2\in\Lambda} \!\!\! \tilde U_{j_{12}} \rho_{\vj_1} \! (\eta) \rho_{\vj_2} \! (\eta) \! \right] \!}
\ee
\normalsize
Here $z = e^{\beta \mu}$ is the activity, $\beta$ is the inverse temperature, $\mu$ is the chemical potential. In this expression $\int \limits_{V} \! (dx)^N = \int \limits_{V} \! dx_1 \ldots \! \int \limits_{V} \! dx_N$ denote integration over the coordinates $x_i = (x_{i}^{(1)},x_{i}^{(2)}, x_{i}^{(3)}) $ of all particles in the system, $\eta = \{ x_1 , \ldots , x_N \}$ is the set of coordinates, $j_{12}= |\vec{j}_{1}- \vec{j}_{2}|$ is the difference between two cell vectors.
 Vectors $\vec{j}_1$ and $\vec{j}_2$ take on values from the set $\Upsilon$, defined as
\[ \label{Lambda}
\Upsilon =\Big\{ \vj = (j_1, j_2, j_3)| j_i = c m_i; m_i=1,2,... N_{a}; \, i=1,2,3;~N_v=N_a^3 \Big\}. \nonumber
\]
Here $c$ is the linear size of each cell, $N_a$ is the number of cells along each axis. Values $\rho_{\vj}(\eta)$ are the occupation numbers of cells~\cite{rebenko_2013,KD_2017,KK_arxiv}.
The interaction potential has the following form
\begin{eqnarray} \label{Morse_pot}
&& \tilde U_{l_{12}}  =  - U_{l_{12}} + \Psi_{l_{12}}, \\
&& \Psi_{l_{12}}  =  D e^{-2(l_{12}-1)/\alpha_R}, \quad U_{l_{12}} = 2 D e^{-(l_{12}-1)/\alpha_R}, \nonumber
\end{eqnarray}
$l_{12}$ is difference between two vectors $\vl_1$ and $\vl_2$ from a set
\[ \label{Lambda}
\Lambda =\Big\{ \vl = (l_1, l_2, l_3)| l_i = c/R_0 m_i; m_i=1,2,... N_{a};\, i=1,2,3;~N_v=N_a^3 \Big\}. \nonumber
\]
moreover $l_{12} = j_{12}/R_0$. $R_0$ corresponds to the minimum of the function $\tilde U_{l_{12}}$ ($\tilde U(l_{12}=1)=-D$ is the depth of potential well). For the point of convenience here and henceforth we measure length in $R_0$-units. Thus $\alpha_R = \alpha / R_0$ is the effective interaction radius $\alpha$ in $R_0$-units.

Looking at \eqref{Morse_pot} it becomes obvious that different particles in the same cell interact with each other equally irrespective of the distance between them. Interaction between constituents of different cells is a function $\tilde U_{l_{12}}$ of distance between cells.

As we had shown in~\cite{KD_2017} in terms of Fourier representation the GPF \eqref{GPF_1} contains a sum of diagonal terms in the exponent. It can be expressed via $N$ integrals over the coordinates of particles and $N_v$ integrals over the collective variables (CV) $\rho_{\vec{k}}$
\small{
\begin{equation}\label{GPF_2}
\Xi = \!\! \sli_{N=0}^{\infty} \! \frac{(z)^N}{N!} \!\! \int \limits_{V} \! (dx)^N \!
\exp \! \left[-\frac{\beta}{2} \!\! \sli_{\vec{k}\in\cB_c} \!\!\! \tilde U(k) \hat \rho_{\vk} \! \hat \rho_{-\vk} \! \right] \int (d \rho)^{N_v} \!\! \int (d \nu)^{N_v} \exp \left[ 2 \pi i \sum \limits_{\vec{k} \in\cB_c} \nu_{\vk} (\rho_{\vk} - \hat \rho_{\vk} )\right] \! .
\end{equation}

}
\normalsize
Herewith
\[ \label{div_ro}
(d\rho)^{N_v} = \prod \limits_{\vk\in\cB_c} d \rho_{\vk}; \quad (d\nu)^{N_v} = \prod \limits_{\vk\in\cB_c} d \nu_{\vk}.
\]
The operator $\hat \rho_{\vk}$ is the representation of the occupation number $\rho_{\vl} (\eta)$ in reciprocal space
\[ \label{ro_site}
\hat \rho_{\vk} = \frac{1}{\sqrt{N_v}} \sli_{\vec{l}\in\Lambda}\rho_{\vl}(\eta) e^{i\vec{k}\vl}.
\]
Vector $\vec{k}$ takes values from the set $\cB_c$ corresponding to one cell
\small{
\[\label{Bk}
\cB_c \! = \! \Big\{  \vk \! = \! (k_{1},k_{2},k_{3}) \Big| k_{i} \! = \! -\frac{\pi}{c}+\frac{2\pi}{c}\frac{n_{i}}{N_{1}}, \, n_{i} \! = \! 1,2,\ldots,N_{a}; \, i=1,2,3; \, N_{v} = N_{a}^{3}   \Big\}. \nonumber
\]
}
\normalsize
The Fourier transform of the Morse potential \eqref{Morse_pot} $\tilde U(k) = -U(k) + \Psi(k)$ ($k = |\vec{k}|$) is as follows
\small
\[ \label{fourier_Morse_pot}
U(k) = U(0) \! \lp 1 +  \alpha_R^2 k^2 \rp^{-2} \!\!\!\!\!\!, \quad \Psi(k) = \Psi(0) \! \lp 1 + \frac{\alpha_R^2 k^2 }{4} \rp^{-2} \!\!\!\!\!\!.
\]
\[ \label{fourier_Morse_pot_0}
U(0) = 16 D \pi\frac{\alpha_R^3}{\upsilon} e^{R_0/\alpha},\quad \Psi(0) = D \pi \frac{\alpha_R^3}{\upsilon} e^{2R_0/\alpha}.
\]
\normalsize
Hence $\chi$ is a real positive parameter $(\chi>0)$, which is fixed for each particular substance. $\upsilon = v/R_0^3$,
$\beta_c=1/k_B T_c$, $k_B$ is the Boltzman constant, $T_c$ is some fixed temperature which will be defined later.
Let us transfer a part of the repulsive interaction $ \chi \Psi(0) >0$ from the initial interaction potential $\tilde U(k)>0$ to the Jacobian of transition from individual coordinates to collective variables in order to write its accurate representation. The similar idea we used in~\cite{KD_2017}.
Now instead of $\tilde U (k)$ we introduce the effective potential of interaction
\be \label{eff_pot}
W(k) = U(k) - \Psi(k) + \chi \Psi(0).
\ee
Easy to see that a sum of $\chi \Psi(0)$ and $- W (k)$ is equal to the initial potential of interaction \eqref{Morse_pot}.

 The difference between \eqref{eff_pot} and analogous expression in~\cite{KD_2016,KD_2017} is that the present explicit expression of effective potential of interaction is temperature-free.
 The GPF of the model in the representation of collective variables $\rho_{\vk}$ has the following form
\begin{equation}\label{GPF_3}
\Xi = \! \int \!\! (d\rho)^{N_v} \! \exp \! \left[ \beta \mu \rho_{0} + \frac{\beta}{2} \! \sum \limits_{\vec{k} \in\cB_c} \!\!\! W(k) \rho_{\vk} \rho_{-\vk} \! \right] \prod \limits_{l=1}^{N_v} \left[ \sli_{m=0}^\infty \frac{v^m}{m!} e^{-pm^2}\delta(\rho_{\vl}-m)\right],
\end{equation}
Note that $\rho_{\vl}$ is the representation of $\rho_{\vk}$ in direct space and $l = | \vl |$ and the parameter $p$ is a function of temperature
\be \label{p1}
p(T) = \chi \beta \Psi (0)/2,
\ee
which is different from analogous temperature-free parameter in~\cite{KD_2016,KD_2017}. This complicates the calculation of the GPF \eqref{GPF_3}.

The second modification of previously developed method~\cite{KD_2016,KDR_2015} is application of Stratonovich-Hubbard transformation to the term which contains the effective potential of interaction
\begin{equation}\label{Strat_Habb_W}
 \exp \! \left[  \frac{\beta}{2} \! \sli_{\vec{k}\in\cB_c} \! W(k) \rhok\rhomk \right] = g_W  \!\!\! \int \!\! (dt)^{N_v} \! \exp  \left[  - \frac{1}{2\beta} \! \sli_{\vec{k}\in\cB_c} \! \frac{ t_{\vk} t_{-\vk}}{W(k)} + \!\! \sli_{\vec{k}\in\cB_c} \! t_{\vk}\rhok \right]
\end{equation}
Note that $W(k) > 0$ for all $\chi > 0$.
\[ \label{gV}
g_W = \prod\limits_{\vec{k}\in\cB_c} \left( 2 \pi \beta W(k)\right)^{-1/2}.
\]
Variables $t_{\vec{k}}$ are complex values $t_{\vec{k}} = t_{\vec{k}}^{(c)} - i t_{\vec{k}}^{(s)}$, for which $t_{\vec{k}}^{(c)}$ and $t_{\vec{k}}^{(s)}$
are real and imaginary parts respectively.

\section{Application of the cumulant representation}

When using the method of collective variables it is convenient to represent the Jacobian of transition $J (\rho_{\vl})$ as a cumulant expansion~\cite{yukhnovskii,KD_2016}
\be \label{J_3}
\tilde  J_l(\tilde t_{\vl}) = \exp \left[ - \sli_{n=0}^\infty \frac{a_n(T)}{n!} \rho_{\vl}^n \right],
\ee
we calculated the functional form of cumulants $a_n(T)$
\small
\bea \label{an}
&&
a_0 (T) = - \ln{T_0 (v, p(T))}; \quad a_1 (T) = - \frac{T_1 (v, p(T))}{T_0 (v, p(T))};  \non
&&
a_2 (T) = - \frac{T_2 (v, p(T))}{T_0 (v, p(T))} + a_1^2; \; a_3 (T) = - \frac{T_3 (v, p(T)}{T_0 (v, p(T))}) - a_1^3(T) + 3 a_1(T) a_2(T); \\
&&
a_4 (T) = - \frac{T_4 (v, p(T))}{ T_0 (v, p(T))} + a_1^4(T) - 6 a_1^2(T) a_2(T) + 4 a_1(T) a_3(T) + 3 a_2^2(T); \nonumber
\eea
\normalsize
However, in contradiction to~\cite{KD_2017} all the cumulants $a_n(T)$ are now functions of temperature since they contain a temperature-dependent parameter $p(T)$ \eqref{p1}. Due to the condition $p(T) > 0$ the special functions $T_n(v, p)$ are rapidly convergent series
\be \label{Tn}
T_n(v, p(T)) = \sli_{m=0}^\infty \frac{v^m}{m!} m^n e^{\frac{-p(T)}{2} m^2}.
\ee

 Taking into account \eqref{Strat_Habb_W} and \eqref{p1} find a precise representation of the GPF of the model
\bea \label{GPF_8}
&&
\Xi  =  g_W e^{-\left(a_0 (T) + \frac{\beta \mu^2}{2W(0)}\right) N_v}   \int (d\tilde t)^{N_v} \exp  \Bigg[  \sqrt{ N_{  v}}  \left(  \frac{\mu}{W(0)} - a_1(T)  \right) \tilde t_0 - \frac{1}{2}  \sli_{\vec{k}\in \cB_c}  D(k) \tilde t_{\vec{k}} \tilde t_{-\vec{k}} - \\
&&
-\sli_{n=3}^\infty \frac{a_n (T)}{n!} N_v^{\frac{2-n}{2}}
\sum_{\substack{\vec{k}_1,...,\vec{k}_n \\ \vec{k}_{i}\in \cB_c}}  \tilde t_{\vec{k}_1}  \ldots \tilde t_{\vec{k}_n}
\delta_{\vk_1+ \ldots +\vk_n}  \Bigg] \nonumber
\eea

where we denote
\[ \label{D_k}
D(k) = a_2 (T) + \frac{1}{W(k) \beta} .
\]

\section{An approximate calculation of the grand partition function}

The expression \eqref{GPF_8} is similar to the one obtained in~\cite{KD_2016,KD_2017} but in present one there is an essential difference. This expression is valid for any values of
\be \label{possible_R0A}
 R_0/\alpha >4 \ln 2
\ee
 as well as arbitrary values of the parameter $\chi > 0$. The inequality \eqref{possible_R0A} is peculiar for description of alkali metals (particularly, $Cs$, $Rb$, $K$, $Na$ and so on) by the Morse potential~\cite{girifalko,lincoln,bringas}.

 It is impossible to calculate \eqref{GPF_8} in the general form, since there is an infinite power series in variable $\tilde t_{\vec{k}}$ in the exponent. In connection with this we use an approximation of $\rho^4$-model consisting in cutting off terms proportional to the fifth power of the variable $\tilde t_{\vec{k}}$ and more ($n_0 \geq 5$).
 In this case \eqref{GPF_8} looks like the functional representation of the 3D Ising model in an external field and respectively belongs to the same universality class~\cite{K_2009}. In our case the chemical potential $\mu$ corresponds to an external field. In order to calculate the GPF \eqref{GPF_8} operate the variable substitution defined by
\[ \label{t_tilde_to_rok}
\tilde t_{\vec{k}} = \rho_{\vec{k}} + a_{34}(T) \sqrt{N_v} \delta_{\vec{k}}, \qquad a_{34}(T) = -\frac{a_3(T)}{a_4(T)},
\]
which is aimed to destroy cubic terms of $\tilde t_{\vec{k}}$. As a result we obtain the following expression
\begin{equation} \label{GPF_9}
\Xi = \! g_W e^{N_v ( E_\mu - a_0(T) )} \!\! \int \!\!  (d\rho)^{N_v} \! \exp \! \Bigg[ \! \sqrt{ \! N_v} M \! \rho_0 - \frac{1}{2} \!\! \sli_{\vec{k}\in\cB_c} \! d(k) \rhok\rhomk - \frac{a_4(T)}{24} \frac{1}{N_v} \! \sum_{\substack{\vec{k}_1,...,\vec{k}_4 \\ \vec{k}_{i}\in \cB_c}} \!\!\! \rho_{\vec{k}_1}...\rho_{\vec{k}_4} \delta_{\vec{k}_1+...+\vec{k}_4}\Bigg],
\end{equation}
with notations
\small
\begin{align}\label{Emu_M_a1r}
& E_\mu = -\frac{\beta W(0)}{2} ( M + \tilde{a}_1(T) )^2 + M a_{34}(T)  + \frac{d(0)}{2}a_{34}^2(T) - \frac{a_4(T)}{24}a_{34}^4(T), \\
& M = \frac{\mu}{W(0)} - \tilde{a}_1(T), \quad \tilde{a}_1(T) = a_1(T) + d(0) a_{34}(T) + \frac{a_4(T)}{6} a_{34}^3(T). \nonumber
\end{align}
\normalsize The coefficient $d(k)$ has the form
\small
\be \label{dk}
d(k) = \frac{1}{\beta W(k)} - \tilde a_2(T), \quad \tilde a_2(T) = \frac{a_4(T)}{2} a_{34}^2(T) - a_2(T).
\ee

\normalsize
On this stage we use a type of mean-field approximation considering only variables $\rho_{\vec k}$ with $\vec k = 0$ (see~\cite{KD_2017}). Applying this approximation one would describe a behavior of the model in wide range of temperature (excluding a narrow vicinity of the critical point where contribution of variables $\rho_{\vec k}$ with $\vec k \neq 0$ is important).

 In this approximation the GPF has the form
\be \label{GPF_10}
\Xi \simeq g'_W  e^{N_v E_\mu} \ili_{-\infty}^\infty d\rho_0 \exp \left[ N_v E(\rho_0)\right].
\ee
$E(\rho_0)$ is obtained using the change of variables $\rho'_0=\sqrt{N_v}\rho_0$
\be \label{Ero}
E(\rho_0)=M \rho_0 - \frac{1}{2} d(0) \rho_0^2 - \frac{a_4(T)}{24} \rho_0^4.
\ee
In the mean-field approximation~\cite{kadanoff} a temperature of transition can be determined from the following condition
\be \label{d0}
d(0) = \frac{1}{\beta_c W(0)} - \tilde{a}_{2c} = 0,
\ee
index $c$ means that value is taken at fixed temperature $T_c$: $a_{nc} = a_n (T_c)$, $W(0) = W(k) \big|_{\vk = 0}$ The expression \eqref{d0} gives a definition of the critical temperature
\be \label{Tc}
k_B T_c = \tilde a_{2c} W(0).
\ee
Taking into account \eqref{eff_pot} it is easy to make sure, that $d(0)$ can be expressed as follows
\begin{eqnarray}\label{d0gam_gam_A0}
 d(0)  &=&   \tilde{a}_{2c} (\tau + 1 ) - \tilde{a}_{2}, \\
 \tau &=& (T-T_c)/T_c   \nonumber
 \end{eqnarray}

 Using the Laplace method~\cite{fedoryuk} we obtain the asymptotic form of GPF as follows
\be \label{GPF_10a}
\Xi \simeq g'_W \exp \left[ N_v  E_\mu + N_v  E(\bar{\rho}_0)\right],
\ee
where the value of $\rho_0 = \bar\rho_0$ corresponds to the maximum of $E(\rho_0)$.
Having an explicit expression of the GPF \eqref{GPF_10a} we can find an equation for average density of the system using a well-known formula
\be \label{aver_N}
\bar n  = \! \frac{1}{N_v } \frac{\partial \ln \Xi}{\partial \beta \mu} = \frac{\partial E_\mu}{\partial\beta\mu}+ \frac{\partial E_0(\bar\rho_0)}{\partial\beta\mu}.
\ee
Taking into account \eqref{aver_N}
\begin{equation}\label{n_bar_ae}
 \bar n \! = \! n_g - M  - \frac{\bar\rho_0}{\beta W(0)},
\end{equation}
here
\be\label{ng}
n_g = - a_1 (T) - a_2(T) a_{34}(T) + \frac{a_4(T)}{3} a_{34}^3(T),
\ee
Taking into account \eqref{Tc} the following equality is obvious
\[ \label{betaW0_ae}
\beta W(0) = \big[ \tilde a_{2c} (\tau + 1)\big]^{-1}
\]
since $\bar\rho_0=\bar\rho_0(\tau,M)$ the expression \eqref{n_bar_ae} is the key expression in a framework of the grand canonical ensemble. The following condition of maximum of $E(\bar \rho_0)$
\be \label{eq_M_TbTc}
M = d(0) \bar\rho_{0} + \frac{a_4(T)}{6} \bar\rho_{0}^3,
\ee
gives an equation bounding the chemical potential $M$ and the density of the system $\bar n$:
\be \label{eq_m_TbTc}
m^3 + m p_b + q_b = 0,
\ee
where $m = M + (\bar n - n_g)$
\[ \label{pbqb}
p_b = \frac{6}{a_4}[\tilde a_{2c} (\tau + 1)]^2 \tilde a_2  \quad q_b = - \frac{6}{a_4}[\tilde a_{2c} (\tau + 1)]^3(\bar n - n_g) ,
\]

\begin{figure}[h!]
\begin{centering}
\includegraphics[width=200pt]{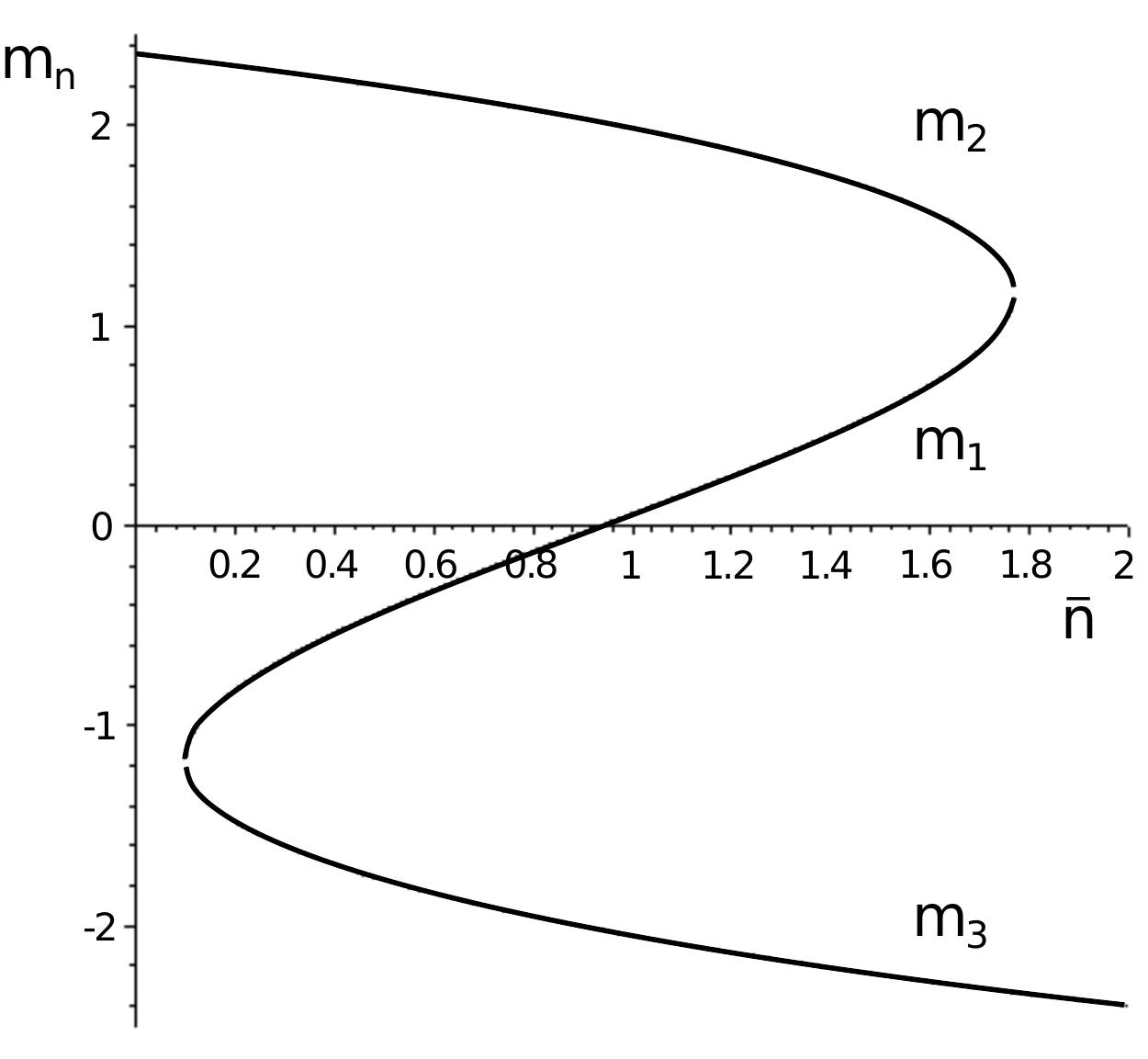}
\vskip-3mm\caption{Plot of solutions $m_n$ as a function of number density $\bar n$ }\label{fig_1}
\end{centering}
\end{figure}

Solutions of this equation are plotted on Figure \ref{fig_1}. Of course, the curve 1 is the one that shows a physical dependence (namely, growth of $m_n$ with increasing density). A solution corresponding to the curve 1 exists on the density interval
\begin{equation}\label{density_interval}
    n_{min} \leqslant \bar n \leqslant n_{max},
\end{equation}
which is compatible to a particular range of values of $m_1$. The chemical potential as a function of density has the following form
\be \label{M_n_bar}
M (\bar n) = m_1 (\bar n) - (\bar n - n_g).
\ee
\begin{eqnarray}\label{m1}
  && m_{1}(\bar n) = 2\sqrt{\frac{2\tilde a_2^3(T)}{a_4(T)}} \sin \frac{\alpha_b(\bar n)}{3} \\
  && \alpha_b(\bar n) = \arcsin \left[ \sqrt{\frac{9 a_4(T)}{8 \tilde{a}_2^3(T)}}(\bar n - n_g)\right]. \nonumber
\end{eqnarray}

Leaning on data of computer experiments~\cite{singh} for sodium and potassium consider $n_{min} = 0.1$. Consequently, there is an equation
\be \label{nmin_0}
n_g = \frac{2}{3} \left[  \frac{ 2 \tilde{a}_2^3(T) }{ a_4 (T) } \right]^{\frac{1}{2}}-0.1,
\ee
by means of which we find the parameter $\upsilon$ as a function of temperature. Change of the density from $0.1$ to the limit value $n_{max}$ is equivalent to the growth of the chemical potential from $M_{min}$ to $M_{max}$.

\begin{figure}[h!]
\begin{centering}
\includegraphics[width=200pt]{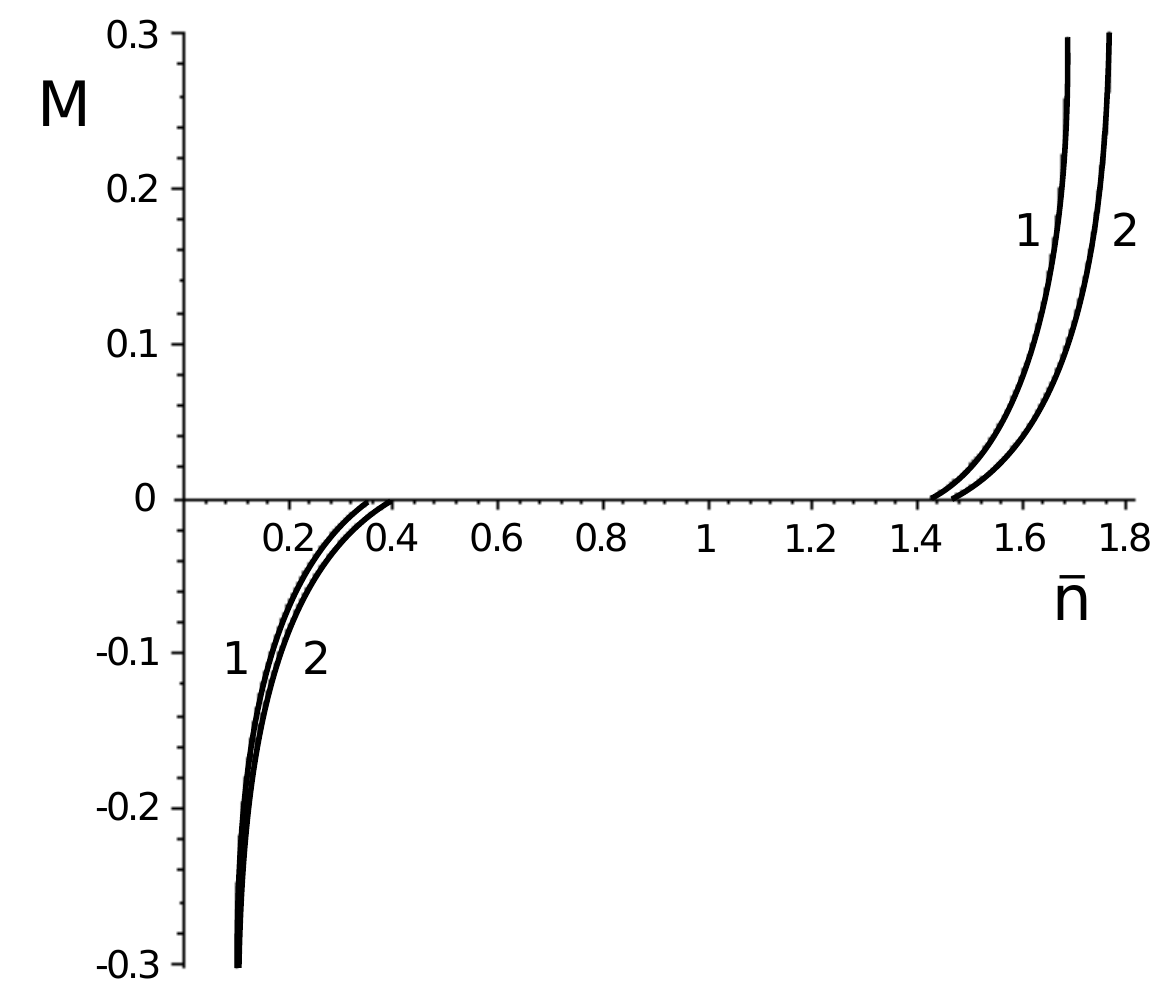}
\vskip-3mm\caption{Plot of the chemical potential $M(\bar n)$ as a function of number density $\bar n$ (curve 1 is for potassium, curve 2 is for sodium). }\label{fig_2}
\end{centering}
\end{figure}

Note that, the solution $m_1$ \eqref{m1}(Figure \ref{fig_1}) is equitable for some (bounded) range of chemical potential values $M=f(\bar n,\tau)$ (see Figure~\ref{fig_2})

\section{Description of the first order phase transitions}

According to the well-known formula $PV = k_B T \ln{\Xi}$ the equation of state of the cell fluid model can be written in the form
\small
\be \label{eq_state_TmTc_1}
\frac{P v}{k_B T} = \frac{\ln g'_W}{N_v} + E_\mu + M(\bar n) \bar\rho_{0i} - \frac{1}{2} \tilde d(0) \bar\rho_{0i}^2 - \frac{a_4}{24} \bar\rho_{0i}^4,
\ee
\normalsize
where $E_\mu$ is defined in \eqref{Emu_M_a1r}, and values $\bar\rho_{0i}$ $i=1,2,3$ are  solutions of equation
\bea \label{ro_eq_TmTc}
&& \bar\rho_{0i}^3 + p_Q \bar\rho_{0i} + q_Q =0, \\
&&  p_Q =  \frac{6 d(0)}{a_4} \qquad q_Q = -\frac{6M(\bar n)}{a_4}. \nonumber
\eea

At $T>T_c$ the discriminant of equation \eqref{ro_eq_TmTc}
\begin{equation}\label{discrQ}
    Q = \left( \frac{2 d(0)}{a_4}\right)^3 + \left( - \frac{3M}{a_4}\right)^2.
\end{equation}
is positive since $p_Q>0$. So we have a single real solution of \eqref{ro_eq_TmTc}. The latter can be found directly from the equation \eqref{ro_eq_TmTc} as follows
\be\label{ro_solution_TbTc}
\bar\rho_{0b} = \lp \frac{3M(\bar n)}{a_4} + \sqrt Q \rp^{\frac{1}{3}} \!\!\!\! + \lp \frac{3M(\bar n)}{a_4} - \sqrt Q \rp^{\frac{1}{3}} \!\!\!\! .
\ee

\begin{figure}[h!]
\begin{centering}
\includegraphics[width=200pt]{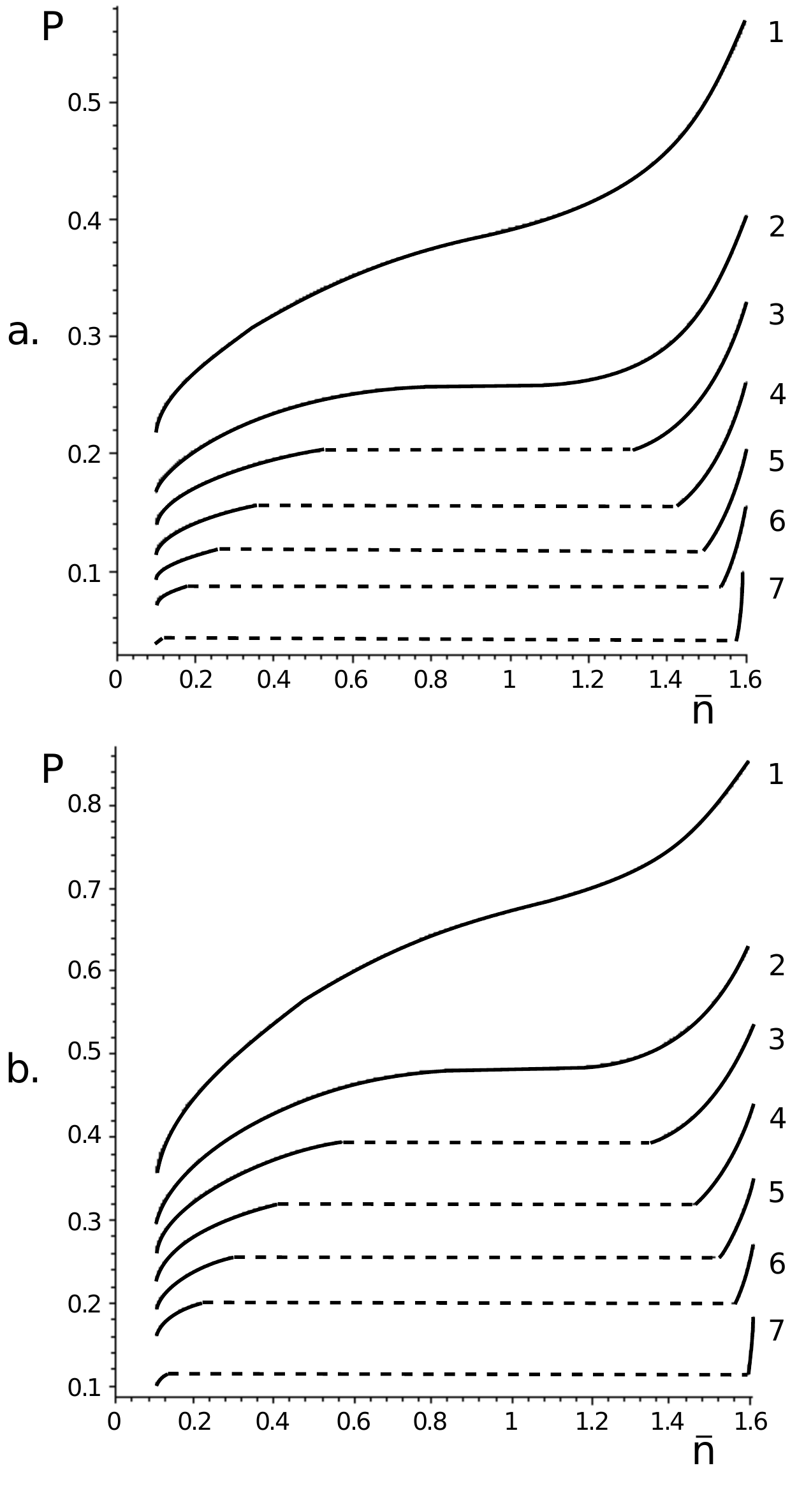}
\vskip-3mm\caption{Plots of the pressure $Р(\bar n)$ as a function of number density $\bar n$ at different temperatures: curve 1 is for $\tau = 0.1$, curve 2 is for $\tau = 0$, curve 3 is for $\tau = - 0.05$, curve 4 is for $\tau = - 0.1$, curve 5 is for $\tau = - 0.15$, curve 6 is for $\tau = - 0.2$, curve 7 is for $\tau = - 0.3$. In figure a. \-- data for potassium, in figure b. \-- data for sodium)}\label{fig_4}
\end{centering}
\end{figure}

 The equation of state or pressure as a function of temperature and density in case of $T>T_c$
\begin{equation} \label{eq_state_TbTc_2}
\frac{P}{k_B T_c} = \frac{\tau + 1}{\upsilon (T)} \big(  f \! + \frac{M (\bar n) }{2 \tilde a_{2c} ( \tau + 1)} \left[ M (\bar n) + 2 \bar n \right]  - \frac{d(0) m_1^2 (\bar n)}{2[\tilde a_{2c} ( \tau + 1)]^2} - \frac{a_4}{24} \frac{m_1^4 (\bar n)}{[\tilde a_{2c} ( \tau + 1)]^4}  \big).
\end{equation}
\small{
\[ \label{f}
f =  \frac{1}{N_v} \ln g'_W - a_0 + \frac{d(0)}{2} a_{34}^2 - \frac{\tilde a_1^2  }{2 \tilde a_{2c} ( \tau + 1)} - \frac{a_4}{24} a_{34}^4.
\]
}
\normalsize An explicit expression of pressure as a function of density at the critical temperature deduced from \eqref{eq_state_TbTc_2} by substituting $T_c$ for $T$ is as follows
\be \label{eq_state_Tc_main}
\frac{P \upsilon|_{T = T_c}}{k_B T_c} = f_c
 + \frac{M_0 (\bar n)}{2 \tilde a_{2c}} \left[ M_0 (\bar n) + 2 \bar n|_{T=T_c} \right] - \frac{a_{4c}}{24} \left [ \frac{m_1 (\bar n)\big |_{T = T_c}}{\tilde a_{2c}} \right]^4,
\ee
\[ \label{fc}
 f_c = \frac{1}{N_v}\ln g'_W - a_{0c} + \frac{a_{4c}}{24} a_{34c}^4 - \frac{\tilde a_{1c}^2}{2 \tilde a_{2c}}.
\]
The expression for total chemical potential $M_{0}$ as a function of density is the following
\[ \label{Mc_solution}
M_{0} =  m_1 (\bar n)\big |_{T = T_c} - (\bar n|_{T=T_c}-n_{gc}).
\]
Indexes $0$ and $c$ denote, that $M_0$ and $a_{nc}$ correspond to the case of $T=T_{c}$.
Plots of pressure dependence $P = P( \bar{n} )$ on average density expressed by \eqref{eq_state_TbTc_2} (curve 1) and $P|_{T=T_c}= P|_{T=T_c}( \bar{n} )$ expressed by \eqref{eq_state_Tc_main} (curve 2) are shown on Figure~\ref{fig_4} for the case of sodium (a) and potassium (b).

\begin{figure}[h!]
\begin{centering}
\includegraphics[width=200pt]{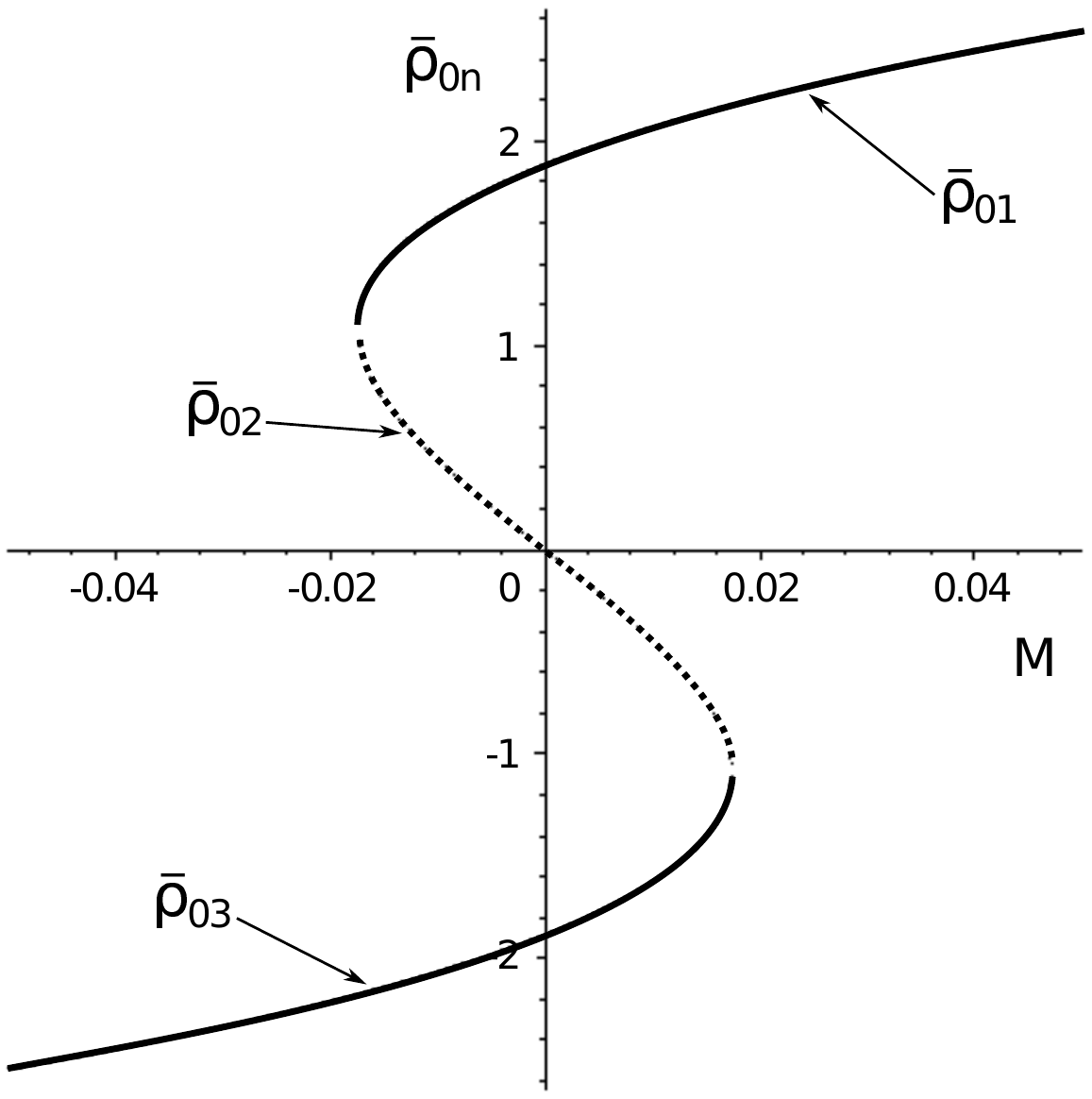}
\vskip-3mm\caption{Plot of solutions $\bar \rho_0n$ as a function of effective chemical potential $M$}\label{fig_3}
\end{centering}
\end{figure}

At $T<T_c$ we have three real solutions of \eqref{ro_eq_TmTc}
\bea \label{solutions_ro123}
&&
\bar\rho_{01} = 2 \rho_{0r}\cos \frac{\alpha_m}{3},\non
&&
\bar\rho_{02} = - 2 \rho_{0r}\cos \lp \frac{\alpha_m+\pi}{3} \rp,\\
&&
\bar\rho_{03} = - 2 \rho_{0r}\cos \lp \frac{\alpha_m-\pi}{3} \rp,\no
\eea
where
\be \label{ror}
\rho_{0r} =  \sqrt{- \frac{2 d(0)}{a_4}},
\ee
and the angle $\alpha_m$
\be \label{Mq}
\alpha_m = \arccos \frac{M}{M_q} \quad M_q = \sqrt{-\frac{8[d(0)]^3}{9 a_4}}.
\ee
The solution $\rho_{01}$ fits the stability condition in the interval $M \in [0,M_{max}]$ as well as $\rho_{03}$ \--- in $M \in [M_{min},0]$ (see Figure \ref{fig_3})

 As a result we can express the equation of state as follows
\bea \label{eq_state_TmTc_4}
&&
\frac{P}{k_B T_c} = - \frac{\tau + 1}{\upsilon (T)} \big(\frac{1}{N_v} ln g'_W + E_\mu(\bar n) + E_1(\bar\rho_{03}) \Theta(\bar n_{12}-\bar n) + E_2(\bar\rho_{03})\Theta(\bar n - \bar n_{12}) \Theta(-\bar n + \bar n_{20}) + \nonumber \\
&&
E_3(\bar\rho_{01}) \Theta(\bar n - \bar n_{03}) \Theta(\bar n_{34} - \bar n) + E_4(\bar\rho_{01}) \Theta(\bar n - \bar n_{34})\big).
\eea
\normalsize The value $E_\mu$ is determined by the formula \eqref{Emu_M_a1r}. Functions $E_n(\bar\rho_0)$ has the following form
\be \label{Enro}
E_n(\bar\rho_{0n}) = M (\bar n)\bar\rho_{0n} - \frac{ d(0)}{2} \bar\rho_{0n}^2 - \frac{a_4}{24} \bar\rho_{0n}^4 ,
\ee
where notation $\bar\rho_{0n}$ is either $\bar\rho_{01}$ from \eqref{solutions_ro123} for $E_3(\bar\rho_{01})$ and $E_4(\bar\rho_{01})$, or $\bar\rho_{03}$ from \eqref{solutions_ro123} for $E_1(\bar\rho_{03})$ and $E_2(\bar\rho_{03})$.
The equation \eqref{eq_state_TmTc_4} also includes values of densities: $\bar n_{12}$ is when $M = - M_q$,
\be \label{n12}
\bar n_{12} = n_g - 2 \tilde a_2 \rho_{0r} + 4 \frac{a_4}{3} \rho^3_{0r},
\ee
$\bar n_{34}$  is when $M = M_q$
\be \label{n34}
\bar n_{34} = n_g + 2 \tilde a_2 \rho_{0r} - 4 \frac{a_4}{3} \rho^3_{0r},
\ee
$\bar n_{20}$ and $\bar n_{03}$ are densities of a liquid-vapor transition
\bea \label{n20}
\bar n_{20} &=& n_g - \sqrt 3 \tilde a_{2c} (\tau + 1) \rho_{0r}, \\
\bar n_{03} &=& n_g + \sqrt 3 \tilde a_{2c} (\tau + 1) \rho_{0r}.
\eea

\section{Analytical results}

\begin{figure}[h!]
\begin{centering}
\includegraphics[width=200pt]{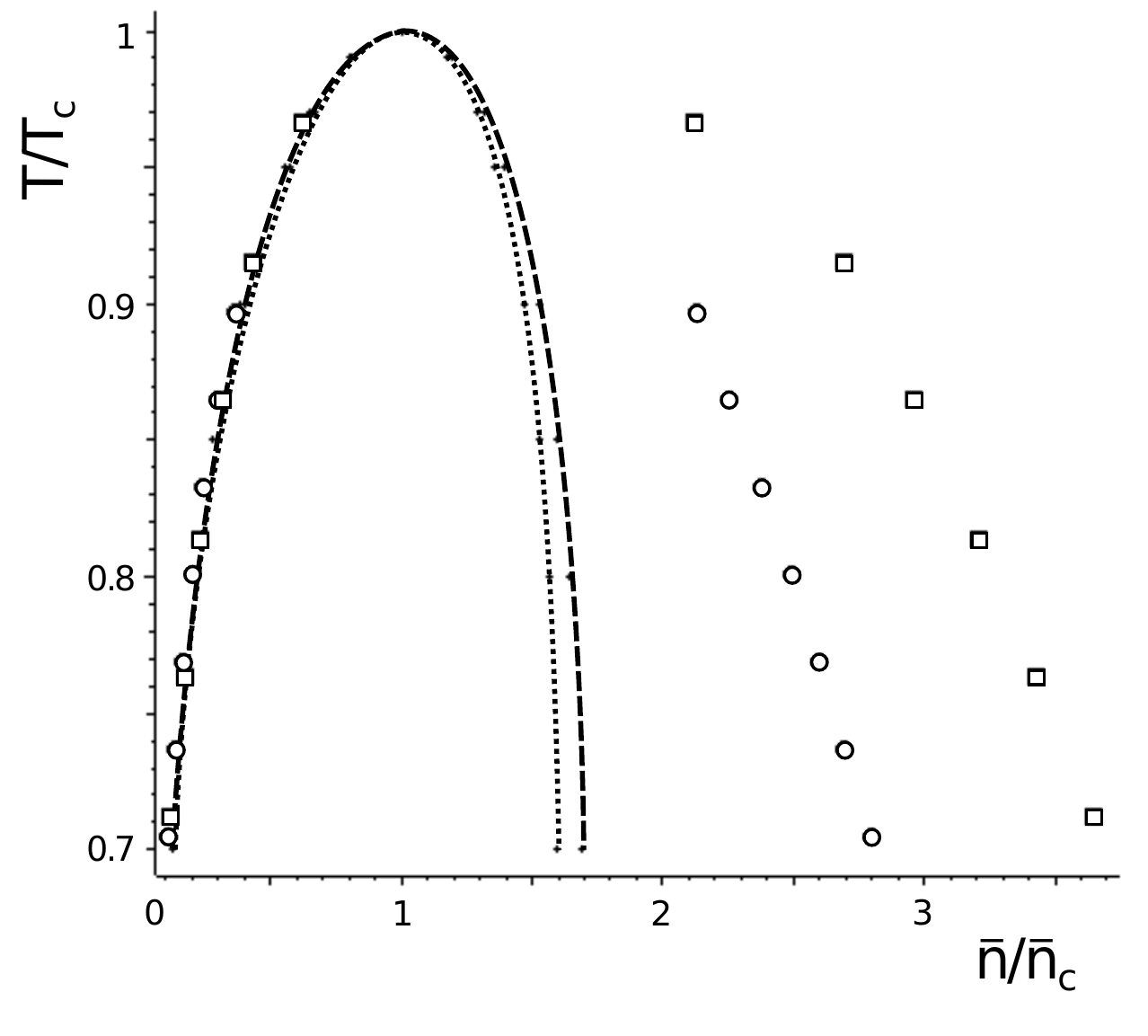}
\vskip-3mm\caption{The coexistence curves: analytical results for K \--- doted curve, Na \--- dashed curve; simulation data~\cite{singh}  K \--- rings, Na \--- boxes, }\label{fig_5}
\end{centering}
\end{figure}
As we mentioned before the liquid-vapor coexistence curves for Na and K has
already been calculated in~\cite{singh} by Monte Carlo simulation in grand
canonical ensemble. Therefore we can compare these with our theoretical results. To do it
we calculated the binodals at the same
temperature interval as in~\cite{singh}. The results of this comparison are
presented in Figure~\ref{fig_5} (using the reduced units $T/T_c$ and $\bar n / \bar n_c$). Both the gas branches of our binodals and these from the simulation data follow the same trend. The agreement is unsatisfying for the liquid branches. The critical point coordinates for sodium and potassium obtained in~\cite{singh} are
\[\label{Na_sim}
    \rho_c^{*} (Na) = 1.430 \qquad T_c^{*}  (Na)= 5.874
\]
\[\label{K_sim}
    \rho_c^{*} (K) = 1.125 \qquad T_c^{*}  (K)= 5.05
\]
 (in reduced units $T^{*} = k_B T/D$ and $\rho^{*} = \rho / R_0^3$). Our results give the following values
\[\label{Na_th}
    \bar n_c (Na) = 0.997 \qquad T_c  (Na)= 5.760
\]
\[\label{K_th}
    \bar n_c (K) = 0.935  \qquad T_c  (K)= 5.037
\]
using the corresponding values of parameters of the model
\[\label{pvx_Na}
  \chi = 1.124 \quad \upsilon \, |_{T_c} = 2.419 \quad \text{for Na},
\]
\[\label{pvx_Na}
\chi = 1.198 \quad \upsilon \,|_{T_c} = 2.940 \quad \text{for K},
\]
according to \eqref{p1} $p (Na) = 1.81$ and $p (K) = 2.01$

As one can see the estimated Na and K critical temperatures are close to the simulations values.
This is however not true for critical densities of both substances, where the analytically obtained critical density
is lower than the value from computer experiments. Never the less in both cases the critical density of sodium is higher than the value for potassium.

\section{Discussion and conclusions}

A theoretical description of the first order phase transition in alkali metals is proposed. Interaction of this type of metals is known to be well described by the Morse potential. The critical density and critical temperature of potassium and sodium is calculated using numeric results for such a potential~\cite{lincoln} both with particular values of microscopic parameters. We obtained a quite good agreement with computer simulation data, despite of applying a type of mean-field approximation. The equation of state is calculated. At the region above the critical temperature isotherms of pressure behave as smooth increasing functions. There is a gas-liquid phase transition below the critical temperature. It is important that in the proposed approach there is no need to use the Maxwell rule. In contradistinction to another approaches connected to the mean-field approximation (for example, the van der Waals theory) a plateau of pressure, which depicts a transition from gas to liquid state, naturally comes of during calculations. This is achieved by applying the Laplace method to calculation of the grand partition function in the $\rho^4$-model approximation. Although the method is approximate we obtained a good agreement with simulation data for the coexistence curves of sodium and potassium in the region of low densities without using any phenomenological parameters.

The introduction of the parameter $\chi$ lies at the heart of the method. It is needed in order to take a certain part of the interaction potential $\chi \Psi (0)$ and use it to calculate the Jacobian of transition from variables in direct space to collective variables. A value of the critical temperature of the model depends on $\chi$. For that reason we choose a value of this parameter so that we obtain values of $T_c$ (for particular substances) which are corespondent to the data of computer experiment. Note that according to the formula \eqref{p1} $\chi$ determines the parameter $p$. The last parameter $\upsilon$ appears as a result of choosing a cell fluid model. Recall that $\upsilon$ is the volume of a cell in $R_0$-units. Due to the self-consistent calculation one gets values of this parameter from the condition \eqref{nmin_0}.

 Plot of binodals on Figure \ref{fig_5} shows that, unfortunately, our approach does not give quantitatively satisfactory results in the fluid region.  As an option, a more complete description can be achieved by introducing phenomenological parameters. Something similar was done by~\cite{bulavin}, the authors obtained good results for fluids with different interaction potentials. On the other hand using approximations of higher power in $\rho$ might be helpful. Taking into account particular results~\cite{nicoll,kulinskii}) we come to the conclusion that appliance of $\rho^m$-models with $m > 4$ stipulate an asymmetry of the coexistence curve in the region of liquid density.

\section*{Acknowledgements}
This work was partly supported by the European Commission under the project STREVCOMS PIRSES-2013-612669, FP7 EU IRSES projects No.612707 (DIONICOS).

\end{document}